\documentclass[12pt,preprint]{aastex}

\usepackage{natbib}
\begin{document}

\def\Msolar{M_{\odot}}

\title{Constraining the properties of the proposed supermassive black 
hole system in 3c66b: Limits from pulsar timing}

\author{Fredrick A. Jenet \altaffilmark{1}, Andrea 
Lommen\altaffilmark{2}, Shane L. Larson\altaffilmark{3}, Linqing Wen 
\altaffilmark{3}}
\altaffiltext{1}{California Institute of Technology, Jet Propulsion 
Laboratory\\ 4800 Oak Grove Drive,Pasadena, CA 91109}
\altaffiltext{2}{Franklin and Marshall College, Department of Physics and Astronomy, PO Box 3003, Lancaster, PA 17604}
\altaffiltext{3}{California Institute of Technology, 1500 
California Blvd. , Pasadena CA, 91125}

\begin{abstract}

Data from long term timing observations of the radio pulsar PSR
B1855+09 have been searched for the signature of gravitational waves
(G-waves) emitted by the proposed supermassive binary black hole
system in 3C66B. For the case of a circular orbit, the emitted G-waves
would generate detectable fluctuations in the pulse arrival times of
PSR B1855+09. General expressions for the expected timing residuals
induced by G-wave emission from a slowly evolving, eccentric, binary
black hole system are derived here for the first time. These waveforms
are used in a Monte-Carlo analysis in order to place limits on the
mass and eccentricity of the proposed black hole system. The reported
analysis also demonstrates several interesting features of a
gravitational wave detector based on pulsar timing.

\end{abstract}


\section{introduction}



This letter reports on the search for gravitational wave (G-wave)
emission from the recently proposed Supermassive Binary Black Hole
(SBBH) system in 3C66B \citep[][S03 hereafter]{sudou03} using 7 years
of timing data from the radio pulsar PSR B1855+09. Given the length of
the available data set and this pulsar's low root-mean-square timing
noise (1.5 $\mu$s), these data are well suited for this analysis. The
proposed binary system has a current period of 1.05 years, a total
mass of $5.4 \times 10^{10} \Msolar$, and a mass ratio of 0.1. Given
the close proximity of the radio galaxy 3C66B (z = 0.02), the G-waves
emitted by this system could induce a detectable signature in the
timing residuals of PSR B1855+09, with a maximum residual amplitude of
order 10 $\mu$s, assuming the eccentricity of the system is zero and
the Hubble constant is 75 km $\mbox{s}^{-1}$ $\mbox{Mpc}^{-1}$

The analysis of these data will demonstrate two interesting properties
of a gravitational wave detector made up of radio pulsars.  First, the
amplitude of the observed signature increases with decreasing
gravitational wave frequency. Second, the light travel time delay
between the Earth and the pulsar can, depending on the geometry, allow
one to observe the gravitational wave source at two distinct epochs of
time simultaneously.  For example, if the pulsar is 4000 light-years
away and the Earth-pulsar line-of-sight is perpendicular to the G-wave
propagation vector, then the observed timing residuals will contain
information about the source both at the current epoch and 4000 years
ago.  If the G-wave emitter is a binary system, slowly inspiraling due to
G-wave emission, then the observed residuals will contain both low and high
frequency components.  The difference in the frequencies of these
components will depend on how quickly the system is evolving.  Since
pulsar timing is more sensitive to lower frequencies, the highest
amplitude oscillations in the timing residuals will be due to the
delayed (i.e. 4000 year old) component.  This effect, referred to as
the ``two-frequency response'', is analogous to the three-pulse response
occurring in spacecraft doppler tracking experiments \citep{ew75} and
the multi-pulse response from time-delay interferometry used in the
proposed Laser Interferometer Space Antenna (LISA) mission
\citep{aetApj}.

The next section describes the expected signature of G-wave emission
from a general binary system and for the specific case of the proposed
system in 3C66B. The observations of PSR B1855+09 used to search for
G-waves are described in section 3.  Section 4 discusses the search
techniques employed as well as the Monte-Carlo simulation used to
place limits on the mass and eccentricity of the system, and the
results are discussed in section 5.

\section{The Signature of 3C66B}

The orbital motion of the proposed binary system in 3C66B will
generate gravitational radiation.  The emitted G-waves will induce
periodic oscillations in the arrival times of individual pulses from
radio pulsars.  Given a model for the pulse arrival times in the
absence of G-waves, one can generate a time series of ``residuals''
which are the observed pulse arrival times minus the expected pulse
arrival times.  Ideally, the effects of known accelerations are
removed from the timing residuals leaving only the variations due to
the presence of G-waves.

The emitted G-waves are described by two functions of spacetime,
$h_+$ and $h_\times$ which correspond to the gravitational wave
strain of the two polarization modes of the radiation field.  As these
waves pass between the Earth and a pulsar, the observed timing
residuals, $R(t)$, will vary as \citep{ew75,Detweiler79}
\begin{equation}
R(t) = \frac{1}{2}(1 + \cos(\mu))(r_+(t) \cos(2 \psi) + r_\times(t) 
\sin(2 \psi)),
\label{dnu}
\end{equation}
where $t$ is time, $\mu$ is the opening angle between the G-wave
source and the pulsar relative to Earth, $\psi$ is
the G-wave polarization angle, and the ``+'' and ``$\times$'' refer to
the two G-wave polarization states.  The functions $r_{+}$ and
$r_{\times}$, referred to collectively as $r_{+,\times}$, are related
to the gravitational wave strain by
\begin{eqnarray}
r_{+,\times}(t) &=& r_{+,\times}^e(t) - r_{+,\times}^p(t) \label{r1}\\
r_{+,\times}^e(t) &=& \int_0^t h_{+,\times}^e(\tau)d\tau \label{r2}\\
r_{+,\times}^p(t) &=& \int_0^t h_{+,\times}^p(\tau - \frac{d}{c}(1 - 
\cos(\mu))) d\tau,\label{r3}
\end{eqnarray}
where $h_{+,\times}^e(t)$ is the gravitational wave strain at Earth,
$h_{+,\times}^p(t)$ is the gravitational wave strain at the pulsar,
$\tau$ is the time integration variable, $d$ is the distance between
Earth and the pulsar, and $c$ is the speed of light. Note that the
pulsar term, $h_{+,\times}^p$, is evaluated at the current time minus
a geometric delay. 


G-waves emitted by a system in a circular orbit (i.e. zero
eccentricity) will vary sinusoidally as a function of time with a
frequency given by twice the orbital frequency. For eccentric systems,
the emitted waves will contain several harmonics of the orbital
frequency. The 2nd harmonic will dominate at low eccentricities while
the fundamental (i.e. the orbital) frequency will dominate at high
eccentricities.  In general, the period and eccentricity of a binary
system will be decreasing with time due to the fact that the system is
radiating away energy and angular momentum in G-waves.  Hence, the
frequencies present in $h_{+,\times}(t)$ will vary with time.  Since
$r_{+,\times}^e$ and $r_{+,\times}^p$ may be generated by
$h_{+,\times}(t)$ at epochs separated by an extremely long time
interval, the frequency content of these terms may differ
significantly.



The G-wave strain, $h(t)$, induced by a black hole binary may be
calculated using the standard weak field approximation applied to two
orbiting point masses \citep{wah87}.  The expected residuals are found
by integrating $h(t)$ with respect to time (see Eqs.\ \ref{r1} -
\ref{r3}):
\begin{eqnarray}
r_+^e(t) &=& \alpha(t)( A(t) \cos(2\phi) - B(t) \sin(2 \phi)) 
\label{w1}\\
r_\times^e(t) &=& \alpha(t)( A(t) \sin(2\phi) + B(t) \cos(2 
\phi)),\label{w2}\\
\alpha(t) &=&  \frac{M_c^{\frac{5}{3}}}{D \omega^{\frac{1}{3}}}\frac{ 
\sqrt{1 - e(t)^2}}{1 + e(t) \cos(\theta(t))} \label{w3}
\end{eqnarray}
where $D$ is the distance to the source,
$\phi$ is the orientation of the line of nodes on the sky, $\omega(t)$
is the orbital frequency, $e(t)$ is the eccentricity, $\theta(t)$
is the orbital phase, and $M_c$ is the ``chirp mass'' defined as
\begin{equation}
M_c = M_t \left(\frac{m_1 m_2}{M_t^2}\right)^\frac{3}{5}
\end{equation}
where $M_t = m_1 + m_2$ and $m_1$ and $m_2$ are the masses of the
individual black holes.  Note that all units from Equation \ref{w1}
on are in ``geometrized'' units where $G=c=1$\footnote{In geometrized
units, mass and distance are in units of time.}. $A(t)$ and $B(t)$ are
given by
\begin{eqnarray}
A(t) &=& 2 e(t) \sin[\theta(t)] \{\cos[\theta(t) - \theta_n]^2 - 
\cos[i]^2 \sin[\theta(t)-\theta_n]^2\} - \nonumber\\
     & & \frac{1}{2} \sin[2(\theta(t)-\theta_n)]\{1 + 
e(t)\cos[\theta(t)]\}\{3 + \cos[2i]\}\\
B(t) &=& 2 \cos[i] \{\cos[2(\theta(t) - \theta_n)] + e(t) 
\cos[\theta(t) - 2\theta_n]\}
\end{eqnarray}
where $i$ and $\theta_n$ are the orbital inclination angle and the 
value of $\theta$ at the line of nodes, respectively \citep{wah87}. 
$\theta(t)$ and $e(t)$ are given by the following coupled 
differential equations \citep{wah87,peters64}:
\begin{eqnarray}
\frac{d \theta}{dt} &=& \omega(t)\frac{\{1 + 
e(t)\cos[\theta(t)]\}^2}{\{1 - e(t)^2\}^{\frac{3}{2}}}\label{w6}\\
\frac{d e }{dt}     &=& - 
\frac{304}{15}M_c^{\frac{5}{3}}\omega_0^{\frac{8}{3}} 
\chi_0\frac{e(t)^{\frac{-29}{19}}[1 - e(t)^2]^{\frac{3}{2}}}{[1 + 
\frac{121}{304}e(t)^2]^{\frac{1181}{2299}}},\label{w7}
\end{eqnarray}
where $\omega_0$ is the initial value of $\omega(t)$ and $\chi_0$ is 
a constant that depends on the initial eccentricity $e_0$:
\begin{equation}
\chi_0 = [1 - e_0^2]e_0^{-\frac{12}{19}}[ 1 + \frac{121}{304}e_0^2 ]^\frac{-870}{2299}.
\end{equation}
$\omega(t)$ is given by
\begin{equation}
\omega(t) = a_0e(t)^{-\frac{18}{19}}[1 - 
e(t)^2]^{\frac{3}{2}}[1 + 
\frac{121}{304}e(t)^2]^{-\frac{1305}{2299}} \label{w8},
\end{equation}
where $a_0$ is determined by the initial condition $\omega(t=0) =
\omega_0$.  The above equations are accurate to first order in $v/c$,
and valid only when both $e(t)$ and $\omega(t)$ vary slowly with time. 
The expressions for $r_{+,\times}^p$ are identical to those for
$r_{+,\times}^e$.  Note that $r_{+,\times}^p$ is evaluated at an
earlier time than $r_{+,\times}^e$ (See Eqs.\ \ref{r2} and \ref{r3}).

\begin{figure}
\plotone{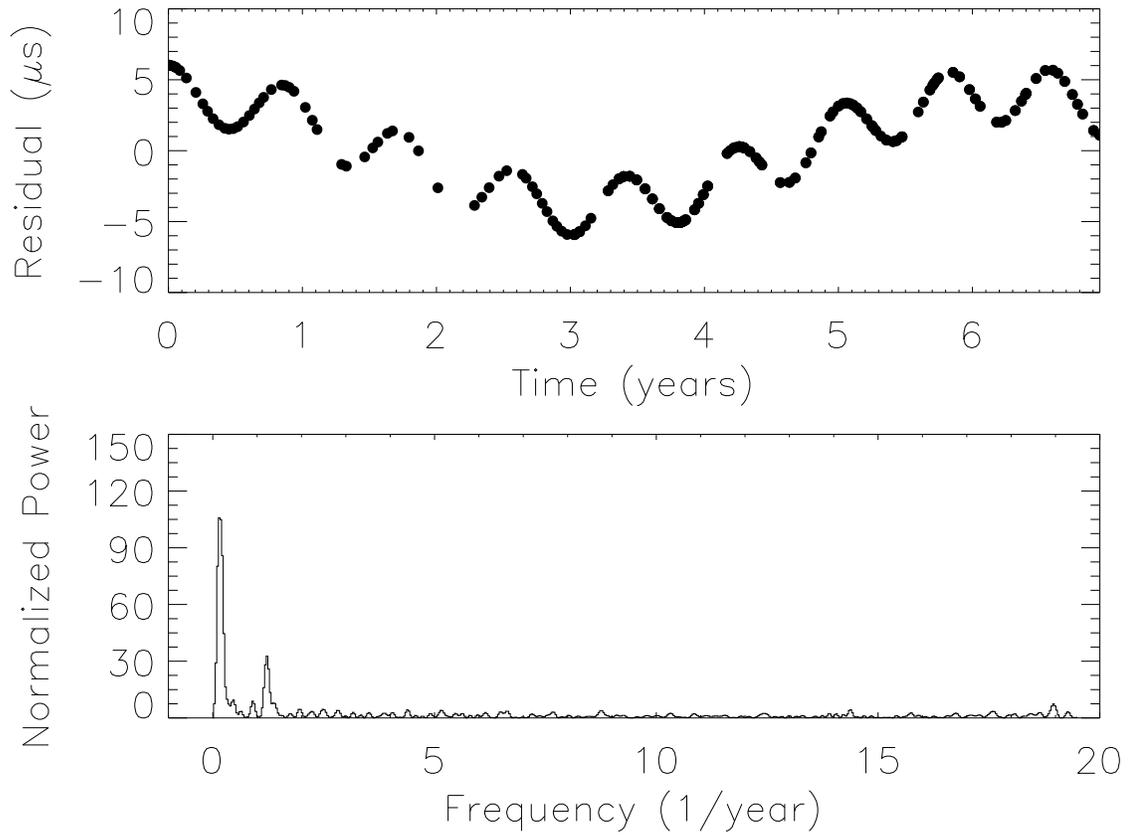}
\caption{Top panel: Theoretical timing residuals induced by G-waves
from 3C66B. The timing points are chosen to coincide with the actual
timing residuals of B1855+09. Bottom panel: The corresponding
normalized Lomb periodogram. }

\label{figure1}
\end{figure}

For the specific case of the S03 parameters for 3C66B, the high chirp
mass ($1.3 \times 10^{10} \Msolar$) together with the period of 1.05
years implies a lifetime of $\approx 5$ years. The orbital period of
such a system will be evolving rapidly. The angle between 3C66B and
PSR B1855+09 on the sky is $81.5^{\circ}$ and PSR B1855+09 lies $1\pm$
0.3 kpc away \citep[][hereafter KTR94]{Kaspi94}. The total time delay
between the pulsar epoch and the Earth epoch is given by $(d/c) (1 -
\cos(\mu))$ which is equal to 3700 $\pm$ 1100 years (see Eq.\
\ref{r3}) for these objects. Since the time delay between the Earth
and the pulsar is much larger than the timescale for evolution of
the system, the expected residual will contain a low frequency
component due to the pulsar term ($r_{+,\times}^p$) and a high
frequency component due to the Earth term ($r_{+,\times}^e$).  The top
panel in Figure \ref{figure1} shows a theoretical set of timing
residuals due to G-wave emission from the proposed binary system in
3C66B assuming that the distance to this galaxy is 80 Mpc and the
distance to the pulsar is 1 kpc. This waveform was generated with $i=
\theta_n = \phi=0$ and $\psi = \pi/4$. The chirp mass used was
$1.3\times 10^{10} \Msolar$ and the orbital period at the epoch of the
S03 observations (i.e. MJD= 51981) was taken to be 1.05 years. The
eccentricity at this epoch was taken to be .0001. Two distinct
oscillation frequencies can be seen, one with a period of about 0.88
years and the other with a period around 6.24 years. The bottom panel
in Figure \ref{figure1} shows the Lomb periodogram of the simulated
residuals. The Lomb periodogram is the analogue of the discrete
Fourier transform for unevenly sampled data and is further discussed
in \S 4. The simulated residuals demonstrate two important features of
an Earth-pulsar gravitational wave detector.  The first is that it is
more sensitive to low frequency oscillations.  The second is that a
single set of timing residuals can contain information about the
source from two widely separated epochs of time.  The low frequency
seen here was due to the orbital period 3700 years ago.

\section{Timing Observations of PSR B1855+09}

We used observations of PSR B1855+09 made by \cite{Kaspi94} (hereafter
KTR94) at the Arecibo Observatory 300 m telescope \footnote[1]{The
National Astronomy and Ionosphere Center Arecibo Observatory is operated
by Cornell University under contract with the National Science Foundation}
and made public therein.  The KTR94 data set is made up of more than 7
years (1986-1993) of bi-weekly observations using the Princeton University
MarkIV system.  The data were corrected for small errors in the
observatory UTC clock as compared to GPS time, for errors in GPS time as
compared to UTC as maintained by the National Institute of Standards and
Technology and finally for errors in UTC(NIST) as compared to terrestrial
time (TT) as maintained by the Bureau International des Poids et Mesures
\citep{Guinot88}. For more details on data acquisition, reduction, and clock
correction, please see KTR94.

\begin{figure}
\plotone{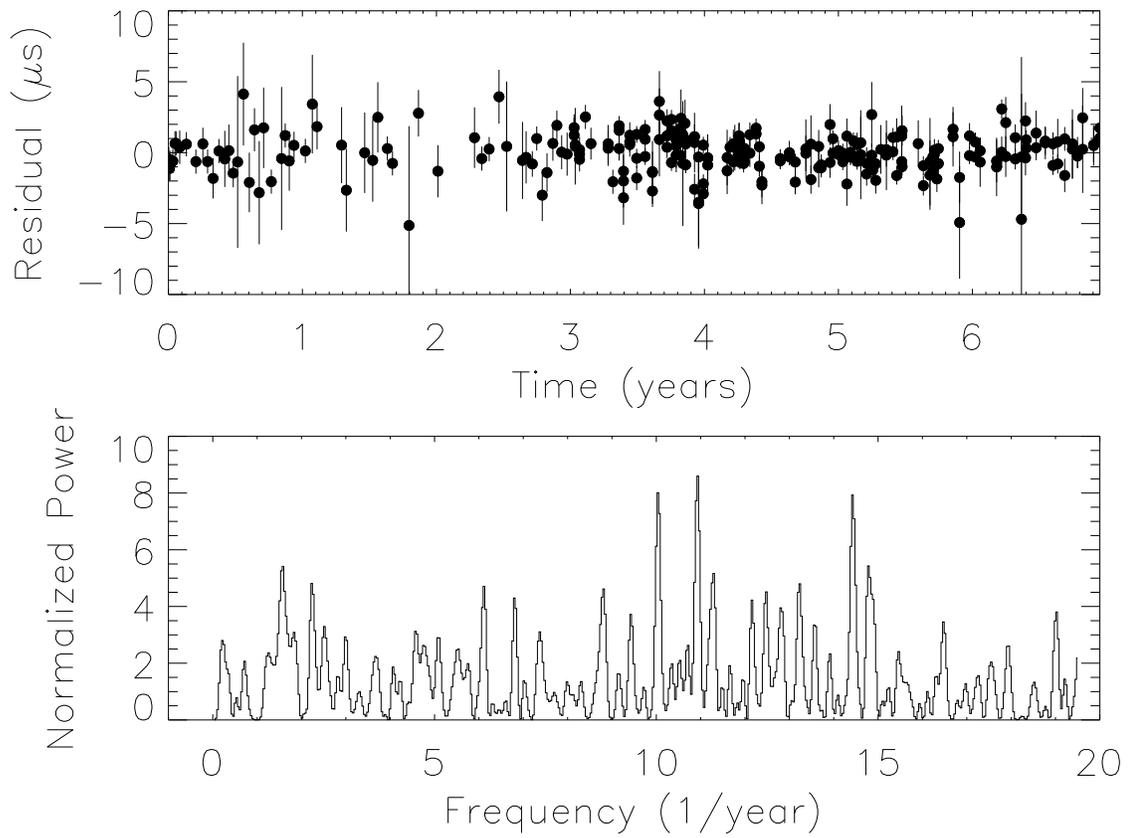}
\caption{Top Panel: Timing residuals for B1855+09. Bottom Panel: The
corresponding normalized Lomb periodogram.}
\label{figure2}
\end{figure}

Using the standard TEMPO software package 
\footnote{http://pulsar.princeton.edu/tempo} the TOAs were fit to the
model published by KTR94. We used their best-fit values as our initial
parameters. In the fitting we allowed the spin period ($P$), period
derivative ($\dot{P}$), right ascension, declination, proper motion,
parallax, and the five Keplerian binary parameters to vary.
Additionally, the Shapiro delay parameters were included in the model
but were fixed at the optimum values published by KTR94.  The best-fit
values of all parameters were consistent with those published by
KTR94. The resulting timing residuals used are shown in the top panel
of Figure \ref{figure2}. 

\section{Constraints from pulsar timing}

The timing residuals from PSR B1855+09 were searched for the signature
of G-waves using the normalized Lomb periodogram (see
\citet{pressetal}, section 13.8) together with ``harmonic summing.''
The Lomb periodogram (LP) is the analog of the discrete Fourier
transform for unevenly sampled data. Harmonic summing is performed by
adding together the periodogram power at harmonics of each frequency
up to a chosen maximum harmonic \citep{lyne88}. This process increases
the sensitivity to periodic, non-sinusoidal waveforms like those
expected from eccentric binaries.  If a SBBH system existed in 3C66B
with an eccentricity of zero and a chirp mass and period adopted by
S03, then the LP should show the two-frequency response like that seen
in Figure \ref{figure1}. Figure \ref{figure2} plots the
normalized LP for the residual data described above.  The periodogram
power was calculated for 542 frequencies ranging from 1/27.8
$\mbox{year}^{-1}$ to 19.5 $\mbox{year}^{-1}$ with a resolution of
1/27.8 $\mbox{year}^{-1}$. This corresponds to a frequency
oversampling factor of 4. There are no significant peaks in this
LP. For purposes of this paper, a significant peak has less than a
$.1\%$ chance of occurring in purely random data assuming Gaussian
statistics. Harmonic summing was performed up to the 6th
harmonic. Again, no significant features were found.

Since the LP analysis was unable to detect the presence of G-waves in
the timing residuals, limits can be placed on the possible chirp mass
and eccentricity of the system.  Since the general waveform given by
Eqs.\ \ref{w1}-\ref{w3} depends on various unknown quantities that
specify the orientation of the orbit and the viewing geometry, a
Monte-Carlo simulation was performed in order to determine the
probability of detecting a SBBH system in 3C66B with a given chirp
mass and eccentricity.  Aside from $M_c$ and $e$, the general wave
form depends on 6 angles: two angles specify the plane of the orbit,
two determine the orbital phase of binary at the beginning of each of
the two relevant epochs, and two determine the initial location of the
line of nodes at the start of each epoch.  For a given $M_c$ and $e$,
the initial eccentricities and periods were determined using Eqs.\
\ref{w6} and \ref{w7}.  An orbital period of 1.05 years at MJD = 51981
was chosen for the initial parameters in order to match the
observations of S03. The distances to 3C66B and B1855+09 were taken to
be 80 Mpc and 1 kpc, respectively. The 6 unknown angles were chosen at
random from a uniform distribution that ranged from 0 to $2\pi$ and a
corresponding waveform was generated using Eqs.  \ref{w1}-\ref{w8}.
The waveform was then added to the residual data. When processing the
timing data, the program TEMPO will remove the effects of the Earth's
orbit and parallax together with a linear trend. In order to simulate
the effects of removing these components from the data, various
functions were subtracted from the simulated data. A one year periodicity
was removed by subtracting a function of the form $y = a \cos(\omega
t) + b \sin(\omega t)$ where $t$ is time, $\omega = 2\pi / 1
~\mbox{year}$, and $a$ and $b$ were determined by a least-squares fit
to the simulated data. A six month periodicity was removed in a similar
fashion. A best fit first order polynomial was also removed. This
combination of data plus simulated signal minus various fitted
functions was then analyzed using the Lomb periodogram method
described above. If a significant peak was found (see above), then the
signal was considered to be detected. 1000 waveforms were tested for
each $M_c$ and initial $e$.  It was found that there is a $98\%$
chance of detecting a system like that adopted by S03 if it has an
eccentricity less than $0.03$.  As the eccentricity increases above
$0.03$, the system is evolving rapidly enough to make the period at
the earlier epoch (i.e. the period in the pulsar term), much longer
than the observation length.  Hence, for eccentricities between $0.03$
and $0.49$, the probability drops to about $95\%$.  The detection
probability starts falling off again above eccentricities of $0.49$.
At this point, the period of binary system at the start of the
observations is longer than the observing time. The results for this
and other chirp masses are summarized in table \ref{table1}.  The
first column lists the chirp mass in $10^{10} \Msolar$, the next four
columns list the limiting eccentricities at the $98\%$, $95\%$, and 
$90\%$ probability levels. For example, with $M_c=1.0$,
if the eccentricity at the epoch of the S03 observations was less than
$0.03$, then there was at least a $95\%$ chance of detecting the
system in these data using the techniques described above.


\begin{deluxetable}{cccccccc}
\tablecaption{\label{table1}Detection Limits}
\tablewidth{4.5in}


\tablehead{\colhead{\scriptsize $M_c$} & \colhead{\scriptsize $98\%$} &
\colhead{\scriptsize $95\%$} & \colhead{\scriptsize $90\%$} &\colhead{\scriptsize $M_c$} & \colhead{\scriptsize $98\%$} & \colhead{\scriptsize $95\%$}
& \colhead{\scriptsize $90\%$} \\ \colhead{\scriptsize $(10^{10} \Msolar)$} & &  & &\colhead{\scriptsize $(10^{10} \Msolar)$} & & & }

\startdata
1.3 & 0.03  & 0.49 & 0.51 & 1.0 &  --   & 0.03 & 0.18  \\
1.2 & 0.02  & 0.49 & 0.51 &0.9 &  --   & 0.02 & 0.04  \\
1.1 & 0.02  & 0.16 & 0.23 &0.8 &  --  &  0.01  & 0.03 \\
    &       &      &      &0.7 &  --  & -- & -- 
\enddata

\tablecomments{{\scriptsize Given a chirp mass, $M_c$, and a minimum
detection probability, this table lists the maximum eccentricity the
proposed system can have at the epoch of the S03 observations. A
``--'' means that the probability of detecting the system never reached the specified value.}}

\end{deluxetable}



\section{Conclusions}

The signature of G-waves emitted by the proposed system in 3C66B was
not found in the analysis of the pulsar timing residuals of PSR
B1855+09.  The system adopted by S03 has a total mass of $5.4 \times
10^{10} \Msolar$ and a chirp mass of $1.3 \times 10^{10} \Msolar$.
The confidence with which such a system can be ruled out depends on
its eccentricity, which is not constrained by the S03 observations.
It is generally accepted that the eccentricity of a system near
coalescence will be small, but exactly how small depends on many
unknown aspects of the system's formation and evolution.  If the
eccentricity is less than $0.03$, then the adopted system may be ruled
out at the $98\%$ confidence level. As the assumed eccentricity of the
system increases, its expected lifetime will decrease. Given that the
system had to exist for longer than one year and assuming that it will
merge when it reaches the last stable orbit, it can be shown that
the eccentricity must be less than 0.3 for a black hole binary system
with negligible spins. In this case, the system can be ruled out at
the $95\%$ confidence level.

Even though the adopted system is highly unlikely, it is possible that
the system has a lower chirp mass. A system with a chirp mass less
than $0.7 \times 10^{10} \Msolar$ cannot be ruled out from the timing
data regardless of the eccentricity. Systems with chirp masses of $1.0
\times 10^{10} \Msolar$ and $0.8 \times 10^{10} \Msolar$ become more
and more allowable when the eccentricities are larger than $0.18$ and
$0.03$, respectively. 

The above discussion assumed a value of 75 km $\mbox{s}^{-1}$
$\mbox{Mpc}^{-1}$ for the Hubble constant. For other values, the chirp
masses listed in Table \ref{table1} need to be multiplied by a factor
of $(H/75)^{-3/5}$ where H is the desired Hubble contant in units of
km $\mbox{s}^{-1}$ $\mbox{Mpc}^{-1}$. For Hubble constants within the
range of 65 to 85 km $\mbox{s}^{-1}$ $\mbox{Mpc}^{-1}$, the chirp
masses listed in Table \ref{table1} are valid to within $10\%$.


Aside from a lower mass binary black hole system, there are other
possible explanations for the S03 observations. The observed
periodicity of 1.05 $\pm 0.03$ years could be an artifact arising
from the Earth's orbit. On the other hand, if the periodicity
is real, then the observed position angles of the two ellipses may be
explained by wandering of the emission region along the jet as various
shocks propagate within the jet (see for example Marscher et al 1991).

This analysis demonstrates how pulsar timing measurements may be used
to search for G-waves from SBBH systems. In the future, pulsar timing
will become more sensitive to SBBH systems as radio astronomers learn
how to reduce the observed noise in pulsar timing data and/or more
stable radio pulsars are discovered. The residual waveforms presented
here will be useful in searching such high quality data for the
signatures of SBBH systems. The two-frequency response may also
provided an interesting tool for studying the physics of such systems
since it will provide information about the SBBH system at two
distinct epochs of time.


Part of this research was performed at the Jet Propulsion Laboratory,
California Institute of Technology, under contract with the National
Aeronautics and Space Administration. AL acknowledges support of NSF
grant 0107342. LW acknowledges support of NSF grants PHY-0071050
and PHY-0107417. The authors wish to thank John Armstrong for useful
discussions.

\end{document}